# INTERNAL GRAVITY WAVES IN A STRATIFIED FLUID WITH SMOOTHLY VARYING BOTTOM

Vitaly V. Bulatov (1), Yuriy V. Vladimirov (1), Vasily A. Vakorin (2)

(1) Institute for Problems in Mechanics, Russian Academy of Sciences, Moscow, Russia.

(2) University of Saskatchewan, Saskatchewan, Canada.

## Abstract

The far field asymptotic of internal waves is constructed for the case when a point source of mass moves in a layer of arbitrarily stratified fluid with slowly varying bottom. The solutions obtained describe the far field both near the wave fronts of each individual mode and away from the wave fronts and are expansions in Airy or Fresnel waves with the argument determined from the solution of the corresponding eikonal equation. The amplitude of the wave field is determined from the energy conservation law along the ray tube. For model distributions of the bottom shape and the stratification describing the typical pattern of the ocean shelf exact analytic expressions are obtained for the rays, and the properties of the phase structure of the wave field are analyzed.

## INTRODUCTION

The propagation of internal gravity waves in stratified natural media (ocean and atmosphere) is affected by the horizontal nonuniformity of the medium. The most characteristic horizontal nonuniformities of the real ocean include variations of the bottom topography, nonuniformity of the density field, and variations of the mean flow. In the present paper the propagation of internal gravity waves of arbitrarily stratified fluid in a layer of varying depth is considered. The difficulty of this problem consist in the fact that when $H=H(x,y)$ ($H$- function of bottom shape) the partial differential equations describing the internal waves do not admit the separations of variables. An exact analytic solution of this problem can be derived, for example by separation of variables, only in the case when the density distribution and the shape of the bottom are described by quite simple model functions [2]. For an arbitrary bottom shape and stratification, it is possible to construct only asymptotic representations of the solution in the near and far regions; however, for describing the internal wave field between these regions an exact numerical solution of the problem is required [2].

If the ocean depth varies slowly as compared with the characteristic length of the internal waves, which is reasonably true for a real ocean, then the method of geometric optics (WKB method) can be used for solving the problem of internal wave propagation above a varying bottom [1,6]. Using the asymptotic representation of the wave field at large distances, from the source [2], we can solve the problem of constructing the uniform asymptotic of the internal waves by a modification of the geometric optics method, namely by the "vertical modes - horizontal rays" method which does not assume that the medium parameters vary slowly with the vertical coordinate[1,3]. In [2] the uniform asymptotic of the far field of the internal waves were constructed for the constant depth case, and it was shown that the far internal waves field is a sum of individual modes each of which is enclosed within its own Mach cone, the asymptotic form of each mode near the corresponding wave front being expressed in terms of certain special function - the Airy function and its derivative, Fresnel integrals [2,5]. Therefore in investigating the problem of internal gravity waves propagation in a stratified fluid with slowly varying bottom must be sought not in the form of usual WKB expansion in harmonic waves but in the form of waves of a special types - Airy and Fresnel waves. In [3] the uniform far field asymptotic were constructed for a source moving in a medium with a density $\rho = \rho(x, y, z)$, and in this case the solution was represented as an expansion in Airy and Fresnel waves.

In the present paper, the problem of the far field of internal waves excited when a source moves in an arbitrarily stratified fluid layer with slowly varying bottom is considered and the uniform asymptotic are found for an isolated mode of the excited wave field.

**1. PROBLEM FORMULATION AND CHOICE OF THE FORM OF THE SOLUTION**

Let a point source of mass move in a layer $-H(\varepsilon x, \varepsilon y) < z < 0$ ($\varepsilon$ is a small parameter) of stratified fluid with the Brunt-Vaisala frequency $N^2(z)$. It is assumed that the velocity $V$ of the source motion along the $x$ axis is greater than the maximum group velocity of the internal waves, that is that the source is moving with supercritical velocity. The linearized hydrodynamic system of equations in the Boussinesq approximation gives [2,4]

$$\frac{\partial^2}{\partial t^2}\left(\Delta + \frac{\partial^2}{\partial z^2}\right)w + N^2(z)\Delta w = \delta''_{tt}(x+Vt)\delta(y)\delta'(z-z_0)$$

$$\Delta \mathbf{u} + \nabla \frac{\partial w}{\partial z} = \delta(z-z_0)\nabla(\delta(x+Vt)\delta(y))$$

$$\Delta = \frac{\partial^2}{\partial x^2} + \frac{\partial^2}{\partial y^2}, \qquad \nabla = \left(\frac{\partial}{\partial x}, \frac{\partial}{\partial y}\right)$$

where $\mathbf{u} = (u_1, u_2)$ is the horizontal velocity vector, $w$ is the vertical velocity component, and $z_0$ is the depth at which the source is moving.

The boundary conditions are the rigid-lid condition at the surface and the impermeability condition at the bottom $z = -H(\varepsilon x, \varepsilon y)$ [2,6]

$$w=0, \quad z=0$$

$$w = \mathbf{u}\nabla H(\varepsilon x, \varepsilon y), \qquad z = -H(\varepsilon x, \varepsilon y) \qquad (1.2)$$

On the basis of the structure of the uniform asymptotic for $H(x,y)=const$ [2] and by analogy with [1,3] the solution of the problem (1.1)—(1.2) will be sought in the form of a sum of modes, each mode propagating independently from the others (adiabatic approximation) and being representable as the following asymptotic series

$$w = A(\varepsilon x, \varepsilon y, z, \varepsilon t) R_0(\sigma) + \varepsilon^a B(\varepsilon x, \varepsilon y, z, \varepsilon t) R_1(\sigma) + \ldots \qquad (1.3)$$

$$u = u_0(\varepsilon x, \varepsilon y, z, \varepsilon t) R_1(\sigma) \varepsilon^{a-1} + \ldots$$

$$R_{i+1}'(\sigma) = R_i(\sigma), \quad \sigma \equiv (S(\varepsilon x, \varepsilon y, \varepsilon t)/a\varepsilon)^a \qquad (1.4)$$

The functions $S(\varepsilon x, \varepsilon y, \varepsilon t)$, $A(\varepsilon x, \varepsilon y, z, \varepsilon t)$, and $u_0(\varepsilon x, \varepsilon y, z, \varepsilon t)$, in (1.3) and (1.4) are to be found. The argument $\sigma(x, y, t)$ is assumed to be of the order of unity. The value of $a$ is equal to *2/3* for a Airy and *a=1/2* for a Fresnel wave. The function $R_0(\sigma)$ is expressed in terms of the Airy function (Airy wave) or Fresnel integrals (Fresnel wave)[2]. If we seek the solution for vertical component of the Airy wave velocity, then as $R_0(\sigma)$ we must take the derivative $dAi(\sigma)/d\sigma$ Airy function

$$Ai(\sigma) = \int_{-\infty}^{+\infty} \exp(-it\sigma - i\frac{t^3}{3}) dt$$

while for the vertical displacement of Airy waves as $R_0(\sigma)$ we must take the Airy function $Ai(\sigma)$. For the vertical component of the Fresnel waves velocity as $R_0(\sigma)$ we will take the function $d\Phi(\sigma)/d\sigma$, and, accordingly, for vertical displacement of the Fresnel wave

$$R_0(\sigma) = \text{Re} \int_0^\infty \exp\left(-it\sigma - i\frac{t^2}{2}\right) dt \equiv \Phi(\sigma),$$

In what follows we will, without loss of generality, consider, for example, the propagation of the vertical displacement $\eta$ ($w = \partial \eta / \partial t$) of the Fresnel wave since by of the linearity of the problem it is possible to go over from the vertical displacement top the vertical velocity by setting $w$ everywhere equal to rise $\partial \eta / \partial t$ [2,3].

## 2. EIKONAL EQUATION AND CHARACTERISTICS. GEOMETRY OF THE PHASE STRUCTURE OF THE WAVE FIELD

Let us now determine the function $S$. Substituting (1.3) and (1.4) in (1.1), with an accuracy of the order of $\varepsilon^{3/2}$ we obtain

$$u_0 = -A_z' \sqrt{2S} \frac{\nabla S}{\varepsilon} \left( \left(\frac{\partial S}{\partial x}\right)^2 + \left(\frac{\partial S}{\partial y}\right)^2 \right)^{-1} + O(\varepsilon^{3/2})$$

Substituting (1.3) and (2.1) in (I.I) and (1.2) and equating the terms with like powers of c gives for $\varepsilon^{1/2}$

$$\frac{\partial^2 A}{\partial z^2} + |k|^2 \left( \frac{N^2(z)}{\omega^2} - 1 \right) A = 0, \qquad (2.2)$$

$$A = 0, \qquad z = 0, -H(x, y),$$

$$k \equiv (p, q) = -\nabla S, \qquad \omega = \partial S / \partial t$$

The dispersion relation denoted henceforth by $K(\omega, x, y)$ is determined from the solution of the vertical spectral problem (2.2), where $\omega$ is the spectral parameter. We then obtain the eikonal equation determining the function $S$

$$(\partial S / \partial x)^2 + (\partial S / \partial y)^2 = K^2(\omega, x, y) \qquad (2.3)$$

Equation (2.3) is the Hamilton-Jacobi equation with the Hamiltonian $|k|^2 - K^2(\omega, x, y)$, and the characteristic system of this equation has the form [1,3]

$$\frac{dx}{d\tau} = \frac{p}{K(\omega,x,y)K'_\omega(\omega,x,y)}, \qquad \frac{dy}{d\tau} = \frac{q}{K(\omega,x,y)K'_\omega(\omega,x,y)}$$

$$\frac{dp}{d\tau} = \frac{K'_x(\omega,x,y)}{K'_\omega(\omega,x,y)}, \qquad \frac{dq}{d\tau} = \frac{K'_y(\omega,x,y)}{K'_\omega(\omega,x,y)}, \qquad \frac{d\omega}{d\tau} = 0 \qquad (2.4)$$

It is convenient to assign the initial conditions for the system (2.4) in the three-dimensional space $(x,y,t)$ on certain surface: $t = t_0$, $x = x_0(l)$, $y = y_0(l)$. Let the eikonal $S$ be known on this surface: $S(x,y,t) = S_0(l,t_0)$. This corresponds to assigning the initial eikonal on a certain fixed line $x = x_0(l)$, $y = y_0(l)$ at an arbitrary moment of time $t = t_0$. Differentiating the initial eikonal with respect to $t_0, l$ we obtain a system of equations for the determining the initial values of the frequency $\omega_0$ and the wave vector $(p_0, q_0)$

$$\omega_0(l, t_0 0 = \frac{\partial S_0}{\partial t_0}$$

$$p_0(l,t_0)\frac{dx_0(l)}{dl} + q_0(l,t_0)\frac{dy_0(l)}{dl} = \frac{\partial S_0}{\partial l}$$

$$p_0^2(l,t_0) + q_0^2(l,t_0) = K^2(\omega_0(l,t_0), x_0(l), y_0(l))$$

Thus, in the three-dimensional space *(x,y,t)* the solution of system (2.4) defines a family of spatial-time rays $x = x(t,t_0,l)$, $y = y(t,t_0,l)$, where $x = x_0(l)$, $y = y_0(l)$ at $t = t_0$. At this case $l$ and the moment of departure of the ray $t_0$ play the part of ray coordinates, and the variable $t$ is simultaneously a Cartesian and ray coordinates. As distinct from the from the characteristic system for a monochromatic wave [1], the projections of the spatial-time rays on the plane *(x,y)* define a two-parameter family of rays, which for fixed $t_0$ goes over into the usual one-parameter family of rays with ray coordinates $t$ and $l$. Nevertheless, in what follows we will call the variables $l$ and $t_0$ the ray coordinates. As may be seen from (2.4), along the ray the frequency $\omega$ preserves its initial values $\omega_0(l,t_0)$ and the eikonal $S^*(t,t_0,l) = S(x(t,t_0,l), y(t,t_0,l),t)$ in ray coordinates is determined by integration along the ray

$$S^*(t,t_0,l) = S_0(l,t_0) + \omega_0(l,t_0)(t - t_0) + \int_{t_0}^{t}\left(\frac{d \ln K(\omega_0(l,t_0),x(\tau,t_0,l),y(\tau,t_0,l))}{d\omega}\right)^{-1} d\tau$$

On order to find *S(x,y,t)* and $\omega(x,y,t)$ in the Cartesian coordinates *x, y, t* it is sufficient to invert the ray equations $l = l(t,x,y)$, $t_0 = t_0(t,x,y)$. For this it is necessary that for any t the Jacobian

$$D(t,t_0,l) \equiv \frac{\partial x(t,t_0,l)}{\partial t_0}\frac{\partial y(t,t_0,l)}{\partial l} - \frac{\partial x(t,t_0,l)}{\partial l}\frac{\partial y(t,t_0,l)}{\partial t_0} \quad \text{be nonzero.}$$

For numerical analysis of internal wave phase structure we will consider the following as a model example admitting analytic solution of the system (2.4) and at the same time giving a qualitatively correct description of the bottom topography. We assume that the Brunt-Vaisaala frequency is constant, $N(z) = N = \text{const}$, and that the depth depends linearly on the $y$ coordinate alone, $H(y) = \beta y$. We introduce the coordinate system with $x$ axis along the shore $y = 0$. We assume that a point mass source is moving from right to left in the negative direction along the $x$ axis with the velocity $V$ at a distance $y_0$ from the shore, and that at any instant $t$ it radiates waves at all frequencies in the range $0 < \omega < N$ We will consider the First mode. Then for $N(z) = \text{const}$ from (2.2) there follows

$$p^2 + q^2 = K^2(\omega, y), \quad K(\omega, y) = \frac{\pi \omega}{H(y)\sqrt{N^2 - \omega^2}}$$

As the function $K(\omega, y)$ does not depend on $x$ from (2.4) we obtain

$$\frac{dp}{d\tau} = 0, \quad p = \frac{\omega}{V} = \text{const}, \quad q = \pm\sqrt{K^2(\omega, y) - \frac{\omega^2}{V^2}}$$

Then the characteristic system and the initial conditions for the eikonal equation have the form

$$\frac{dx}{d\tau} = \frac{\alpha^2 \gamma^2}{V^2} y^2, \quad \frac{dy}{d\tau} = \pm \alpha^{3/2} \gamma\, y \sqrt{1 - \frac{\alpha \gamma^2 y^2}{V^2}}$$

$$x = V t_0, \quad t = t_0, \quad y = y_0, \quad t = t_0, \qquad (2.5)$$

$$\alpha = 1 - \frac{\omega^2}{N^2}, \quad \gamma = \frac{N\beta}{\pi}$$

Here and below, the upper sign corresponds to the region $y > y_0$ and the lower sign to the region $y < y_0$. Integrating the system (2.5) gives the equations for the rays

$$y = \frac{V}{\gamma \sqrt{\alpha}}\left[\text{ch}\left(\pm \text{arch}\left(\frac{V}{\gamma \sqrt{\alpha}\, y_0}\right) - \gamma \alpha^{3/2}(t - t_0)\right)\right]^{-1}$$

$$x = x_0 \pm y_0 \sqrt{\frac{V^2}{\alpha \gamma^2 y_0^2} - 1} - \frac{V}{\gamma \sqrt{\alpha}} \text{th}\left(\pm \text{arch}\left(\frac{V}{\gamma \sqrt{\alpha}\, y_0}\right) - \gamma \alpha^{3/2}(t - t_0)\right) \qquad (2.6)$$

$$x = x_0 + \frac{\gamma \sqrt{\alpha}}{V} y\, y_0\, \text{sh}\left(\gamma\, \alpha^{3/2}(t - t_0)\right)$$

As follows from (2.4), the rays (2.6) are curves in the space $x, y$ along which the frequency $\omega$ is constant. Since the group velocity of propagation of the internal waves is maximal at $\omega = 0$ [2,4] ($\alpha = 1$), the position of the wave front with respect to the coordinate system moving with the source ($\xi = x + V t$) is determined from the equation

$$\frac{d\xi}{dy} = \pm \frac{\sqrt{V^2 - (\gamma y)^2}}{\gamma y}, \quad \xi(y_0) = 0 \qquad (2.7)$$

The solution of (2.7) has the form

$$\xi = \pm \frac{V}{\gamma}(d_1(y) - d_2(y))$$

$$d_1(y) = \text{arch}\left(\frac{V}{\gamma y_0}\right) - \text{arch}\left(\frac{V}{\gamma y}\right)$$

$$d_2(y) = \sqrt{1 - \left(\frac{\gamma y_0}{V}\right)^2} - \sqrt{1 - \left(\frac{\gamma y}{V}\right)^2}$$

The eikonal $S^*$ has the form

$$S^* = \omega(t - t_0) + \int_{t_0}^{t} \frac{K(\omega, y(\tau, t_0, \omega))}{K'_\omega(\omega, y(\tau, t_0, \omega))} d\tau = \omega(t - t_0) - \omega\,\alpha\,(t - t_0) = \frac{\omega^3}{N^2}(t - t_0)$$

Then, in the coordinate system moving with the source the curves of constant phase, that is curves along which the eikonal $S^*$ is constant, have the form

$$y = \frac{V}{\gamma \sqrt{\alpha}} \, \text{ch}\left[\left(\pm \, \text{arch}\left(\frac{V}{\gamma \sqrt{\alpha} \, y_0}\right) - \gamma \, \alpha^{3/2} T\right)\right]^{-1} \qquad (2.8)$$

$$\xi = V T - \frac{\gamma \sqrt{\alpha}}{V} y \, y_0 \, \text{sh}(\gamma \, \alpha^{3/2} T), \quad T = \frac{S^* N^2}{\omega^3}$$

The Figures A,B show the calculation results for the following parameters, which are close to the parameters of a real ocean shelf : $\beta = 0.1$, $N = 0.01 \, s^{-1}$, $Y_0 = 1000 \, m$, and $V = 2 N \beta \, y_0 / \pi$ (A), $V = 3 N \beta \, y_0 / \pi$ (B). The ray family envelope or caustic curve shown in the figure is the locus of points $\xi = \xi(y)$ satisfying both (2.8) and the condition $\frac{d\xi}{dy} = 0$. Therefore, Fresnel wave with a given value of the phase $S^*$ has corresponding critical values in the variable $y$ bounding its region of propagation from the traverse of the source motion. From the results presented it can be seen that as the phase $S^*$ increases the region of ray penetration (with respect to $y$) increases.

Thus, an increase in the distance $y$ from the traverse of the source motion leads in the incoming wave field to a decrease in the share of the low-frequency components, that is the Fresnel waves with small phase values (the group velocity decreases with increase in frequency) and with the chosen relation for the bottom topography (the group velocity increases with increase in the distance from the shore $y = 0$), and in view of this the relatively high-frequency components of the field with a lower group velocity may propagate to larger distances $y$ from the shore than the low-frequency components.

## 3. ENERGY CONSERVATION AND AMPLITUDE DETERMINATION

In order to determine the amplitude $A$ we substitute (1.3) and (1.4) in (1.1) and (1.2) and equate the terms of the order of $\varepsilon^{3/2}$. As a result, using the properties of the Fresnel integral, we obtain

$$\omega^2 B''_{zz} + |k|^2 \left( N^2(z) - \omega^2 \right) B = \left( N^2(z) - \omega^2 \right)\left( 2\nabla S \nabla^* A + A \Delta S \right) +$$

$$2\omega \left( \frac{\partial^*}{\partial t} A''_{zz} - |k|^2 \frac{\partial^* A}{\partial t} \right) + \frac{\partial^* \omega}{\partial t} \left( A''_{zz} - |k|^2 A \right) - 4\omega A \nabla S \nabla^* \omega \qquad (3.1)$$

$$B = 0, \quad z = 0$$

$$B = 2\nabla S \nabla H\, A'_z \sqrt{S} \left( (\partial S / \partial x)^2 + (\partial S / \partial y)^2 \right)^{-1}, \quad z = -H(x,y) \qquad (3.2)$$

$$\nabla^* = \nabla + \frac{\partial}{\partial \omega} \nabla \omega \quad,\quad \frac{\partial^*}{\partial t} = \frac{\partial}{\partial t} + \frac{\partial}{\partial \omega} \frac{\partial \omega}{\partial t}$$

We then express the function $A(x,y,z,t)$ in the form

$$A(x,y,z,t) = \psi(x,y,\omega(x,y,t))\, f(x,y,z,\omega(x,y,t))$$

where f is a normalized eigenfunction of the problem (2.2)

$$\int_{-H(x,y)}^{0} \left(N^2(z) - \omega^2\right) f^2(x,y,z,\omega) \, dz = 1$$

We then consider equation (3.1) along the characteristics (2.4); in this case the function $|k| = K(\omega, x, y)$ is assumed to be known. We multiply (3.1) by $A$ and integrate it with respect to $z$ from $-H(x,y)$ to zero. As a result, after some quite cumbersome transformations we obtain

$$\psi^2 \frac{\partial^*}{\partial t}(K K'_\omega) + K K''_\omega \frac{\partial^2 \psi^2}{\partial t} + 2\omega K'_\omega \psi^2 \frac{\partial^*}{\partial t}\left(\frac{K}{\omega}\right) + 2\frac{\omega}{K} \psi^2 \nabla S \nabla^*\left(\frac{K}{\omega}\right) +$$

$$\psi^2 \Delta S + \nabla S \nabla^* \psi^2 = 2 \nabla H \nabla S \frac{\omega^2}{K^2}\left(A'_z(x,y,-H(x,y),t)\right)^2 \qquad (3.3)$$

We next use the horizontal properties of the problem (2.2) assuming that $\omega$ is a fixed spectral parameter. Applying the operator $\nabla$ to (2.2) gives

$$(\nabla A)''_{zz} + K^2(\omega, x, y)\left(\frac{N^2(z)}{\omega^2} - 1\right)\nabla A + A \nabla\left(K^2(\omega, x, y)\left(\frac{N^2(z)}{\omega^2} - 1\right)\right) = 0 \qquad (3.4)$$

We multiply (3.4) by $A$ and integrate with respect to $z$ from $-H(x,y)$ to zero. Then, taking (3.2) into account we obtain

$$\left(A'_z(x,y,-H(x,y),t)\right)^2 \nabla H(x,y) = -\nabla\left(\frac{K^2}{\omega^2}\right) \int_{-H(x,y)}^{0} (N^2(z) - \omega^2) A^2(x,y,z,t) \, dz = -\nabla\left(\frac{K^2}{\omega^2}\right) \psi^2$$

As a result, the expression (3.3) can be represented in the form

$$\frac{d\ln\psi^2}{dt} + \frac{d}{dt}\ln\frac{K^2}{\omega^2} - 2\frac{\nabla S \, \nabla \ln(K^2\omega^{-2})}{K K'_\omega} - \frac{\partial^*}{\partial t}\ln(K K'_\omega) - \frac{\Delta S}{K K'_\omega} = 0$$

$$\frac{d}{dt} = \frac{\partial^*}{\partial t} + \frac{\nabla S}{K K'_\omega}\nabla^*$$

Here, $d/dt$ is a derivative along the characteristics of (2.4). Since the frequency $\omega$ is conserved along the characteristics, we have

$$\frac{d}{dt}\ln\left(\frac{K^2}{\omega^2}\right) = \frac{\nabla S \, \nabla \ln(K^2\omega^{-2})}{K K'_\omega}$$

From this we obtain

$$\frac{d}{dt}\ln\left(\frac{\psi^2}{K^2}\right) + \frac{\Delta S}{K K'_\omega} + \frac{\partial^*}{\partial t}\ln(K K'_\omega) = 0$$

Further, we have

$$\frac{\Delta S}{K K'_\omega} + \frac{\partial^*}{\partial t}\ln(K K'_\omega) = \frac{\Delta S}{K K'_\omega} + \frac{\partial^*}{\partial t}\ln(K K'_\omega) + \frac{\nabla S}{K K'_\omega}\nabla^*\ln(K K'_\omega) - \frac{\nabla S}{K K'_\omega}\nabla^*\ln(K K'_\omega) =$$
$$\frac{d}{dt}\ln(K K'_\omega) + \frac{\Delta S}{K K'_\omega} - \frac{\nabla S}{K K'_\omega}\nabla^*\ln(K K'_\omega) = \frac{d}{dt}\ln(K K'_\omega) + \nabla^*c, \qquad c = -\frac{\nabla S}{K K'_\omega}$$

where $c$ is the vector of the group velocity of internal wave propagation. Then, taking into account that along the characteristics (2.4) $\nabla^* c = \operatorname{div} c$, we finally obtain

$$\frac{d}{dt}\ln\left(\frac{\psi^2 K'_\omega}{K}\right) + \operatorname{div} c = 0 \qquad (3.5)$$

In accordance with the Liouville theorem [1,3], equation (3.5) can be rewritten in the form

$$\frac{d}{dt}\ln(D\psi^2 K'_\omega K^{-1}) = 0 \qquad (3.6)$$

Here, $D$ is the Jacobian of the transformation from the ray coordinates to the Cartesian coordinates. It should be noted that the conservation law (3.6) derived can be interpreted as the energy conservation along the ray tube, in contrast to the case of source motion in a stratified medium with the density $\rho = \rho(x,y,z)$ [3]. In fact, from the averaged hydrodynamic equations it follows that if the non-perturbed density is a function of the horizontal coordinates, then from the existence of a steady density distribution $\rho = \rho(x,y,z)$ the existence of steady flows will follow. However, as these flows are slow, they can be neglected in the first approximation. For this reason $\rho = \rho(x,y,z)$ is usually assumed to be a certain background density field formed as a result of the action of a body force and non-adiabatic sources and is usually prescribed *a priori*, for example, by experiment. Since the medium is not at equilibrium when $\rho = \rho(x,y,z)$ the energy flux is not constant along the ray tube. However, the system with a source and uneven bottom considered in the present study is conservative, there is no energy flux from outside, and therefore the law (3.6) is the law of wave energy conservation along the ray tube.

**4. RAY AND AMPLITUDE SOLUTION STRUCTURE IN A LAYER OF CONSTANT DEPTH**

We will now describe the ray and amplitude structure of the solution for a source moving in a stratified fluid layer of constant depth, which will later make it possible to construct the uniform asymptotic of an isolated mode. We assume that a point mass source moving in the negative direction along the $x$ axis at the velocity $V$ passes through the coordinate origin at the time $t=0$. At each instant the source radiates waves at all frequencies in the range $0 < \omega < \max_z N(z)$. The frequency $\omega$ is constant on a ray directed along the vector $k(p,q)$, where

$$p = \frac{\omega}{V}, \quad q = \left(K^2(\omega) - \omega^2 V^{-2}\right)^{1/2} \equiv \lambda(\omega)$$

Next we write the equations of the rays. In doing so, it is convenient to use the frequency $\omega$ and the instant $t_0$ a ray leaves the source as the ray coordinates. In the case under consideration $(H(x,y) = const)$ the rays are straight lines

$$x = -Vt_0 - \frac{\omega(t-t_0)}{VK(\omega)K'(\omega)}, \quad y = \frac{\lambda(\omega)(t-t_0)}{K(\omega)K'(\omega)} \qquad (4.1)$$

Using these equations, we can write for the Jacobian $D$ and eikonal $S^*$ in the ray coordinates

$$D(t, t_0, \omega) = D(t - t_0, \omega) = V \frac{\partial^2 \lambda(\omega)}{\partial \omega^2} \left(\frac{\lambda(\omega)}{K(\omega)K'(\omega)}\right)^2 (t - t_0)$$

$$S(x,y,t) = S^*(t, t_0, \omega) = S^*(t - t_0, \omega) = \left(\omega - \frac{K(\omega)}{K'(\omega)}\right)(t - t_0)$$

It must be stressed that in a coordinate system moving together with the source, a family of rays is one-parametric with a parameter $\omega$ and represents a fan of rays issuing from the source and confined within an angle $2\tan^{-1}\left[c_0/(V^2 - c_0^2)^{1/2}\right]$, where $c_0 = \left(\frac{dK(\omega)}{d\omega}\right)^{-1}$, $\omega = 0$. The equations of the rays in the coordinate system have the form

$$\xi = \frac{\omega(t - t_0)}{V K(\omega) K'(\omega)}, \quad y = \frac{\lambda(\omega)(t - t_0)}{K(\omega) K'(\omega)}$$

where $\xi = x + Vt$ is the distance from the source to the observation point measured along the $x$ axis in the moving coordinate system.

The expression for the eikonal $S^*$ in the moving coordinate system has the form

$$S^* = \frac{\omega}{V}\xi - \lambda(\omega) y \equiv \mu(\lambda)\xi - \lambda y$$

For a source moving in a stratified layer of constant depth the uniform asymptotic of the elevation $\eta$ have the form [2]

$$\eta = Q(\xi, \lambda, z, z_0)\Phi(\sigma), \quad \sigma = \sqrt{2\xi(\mu(\lambda) - \mu'(\lambda)\lambda)}$$

$$Q(\xi, \lambda, z, z_0) = \frac{V \mu^2(\lambda)\varphi(z,\lambda)}{\sqrt{2\mu''(\lambda)\xi}\,(\mu^2(\lambda) + \lambda^2)} \frac{\partial \varphi(z,\lambda)}{\partial z_0}$$

$$\int_{-H}^{0} N^2(z)\varphi^2(z,\lambda)\,dz = 1, \quad \varphi^2 = f^2 \frac{K(\omega)}{\omega K'(\omega)}$$

Then, using the following equalities

$$\mu'(\lambda) = \frac{1}{V\lambda'(\omega)}, \quad \mu''(\lambda) = -\frac{\lambda''(\omega)}{V(\lambda'(\omega))^3},$$

$$K'(\omega) = \frac{\lambda(\omega)\lambda'(\omega) + \omega V^{-2}}{K(\omega)}$$

we obtain the expression for $\eta$ in the ray coordinates

$$\eta(t,t_0,\omega) = Q^*(t,t_0,\omega)\Phi(\sigma^*), \quad \sigma^* = \sqrt{2S^*(t,t_0,\omega)} \qquad (4.3)$$

$$Q^*(t,t_0,\omega) = \frac{\omega f(z,\omega)\sqrt{K(\omega)K'(\omega)}}{V\lambda(\omega)\sqrt{2|\lambda''(\omega)|\lambda(\omega)(t-t_0)}}\frac{\partial f(z_0,\omega)}{\partial z_0}$$

This gives an explicit expression for the amplitude $\psi$ in the case of constant depth

$$\psi = \frac{\omega\sqrt{K(\omega)K'(\omega)}}{V\lambda(\omega)\sqrt{2|\lambda''(\omega)|\lambda(\omega)(t-t_0)}}\frac{\partial f(z_0,\omega)}{\partial z_0} \qquad (4.4)$$

## 5. UNIFORM ASYMPTOTICS OF AN ISOLATED MODE

In the. ray coordinates the conservation law (3.6) can be rewritten in the form

$$D(t,t_0,l)\psi^2(t,t_0,l)\frac{K'_\omega(t,t_0,l)}{K(t,t_0,l)} = E(t_0,l) \qquad (5.1)$$

where $E(t_0,l)$ is a certain function which is determined by the particular form of the problem in question and can be found using the solution of the problem of the motion of a point mass source in a stratified fluid layer of constant depth. To do this, we use the localization principle, that is we will assume that at typical distances at which the uniform asymptotic (4.3) hold true in a layer of constant depth (of the order of several times the layer thickness [2]) the bottom depth can be assumed to be locally constant. Therefore, at these distances the stratified fluid layer has a locally-constant depth. In this case the function on the right side of (5.1) can be calculated assuming the depth to be locally constant and using frozen horizontal parameters.

At the time $t = t_0$ let the moving source be at the point $x_0, y_0, z_0$. Then, by using (4.2) and (4.4), it is possible to obtain

$$E(t_0,\omega) = \frac{\omega^2}{2V\lambda(\omega,x_0,y_0)K^2(\omega,x_0,y_0)}\left(\frac{\partial f(\omega,x_0,y_0)}{\partial z_0}\right)^2$$

As a result, for a source moving in a stratified fluid layer with slowly varying bottom the first term of the uniform asymptotic of the elevation of the Fresnel wave has the form

$$\eta = \eta_0 \Phi(\sigma^*) f(x,y,z,\omega), \quad \sigma^* = \sqrt{2S(t,t_0,\omega)}$$

$$\eta_0(t,t_0,\omega) = \omega \frac{\partial f(x_0,y_0,z_0,\omega)}{\partial z_0} \left(2V\lambda(\omega,x_0,y_0)K^2(\omega,x_0,y_0)\right)^{-1/2} \left(\frac{K(\omega,x,y)}{K'_\omega(\omega,x,y)D(t,t_0,\omega)}\right)^{1/2}$$

$$x = x(t,t_0,\omega), \quad y = y(t,t_0,\omega)$$

Therefore, the solution obtained describes in very general form the uniform asymptotic of the far field of the internal gravity waves for a source moving above a slowly varying bottom, since for $H = (x,y) = const$ this solution coincides with the uniform asymptotic [2]. For large values of $\sigma^*$ (far from the wave front), by using the asymptotic behavior of the Fresnel integrals for large arguments [5], it is possible to obtain the representation of the solution in the form of an expansion in locally-harmonic waves - usual WKB expansion. For small values of $\sigma^*$ the solutions obtained describe the asymptotic of the field in the vicinity of the wave front of an isolated mode.

The research described in this publication was made possible in part by the Russian Foundation for Basic Research, Grant 99-01-00856.

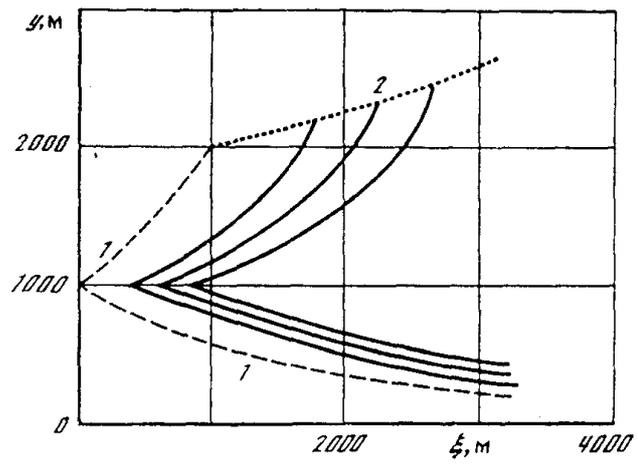

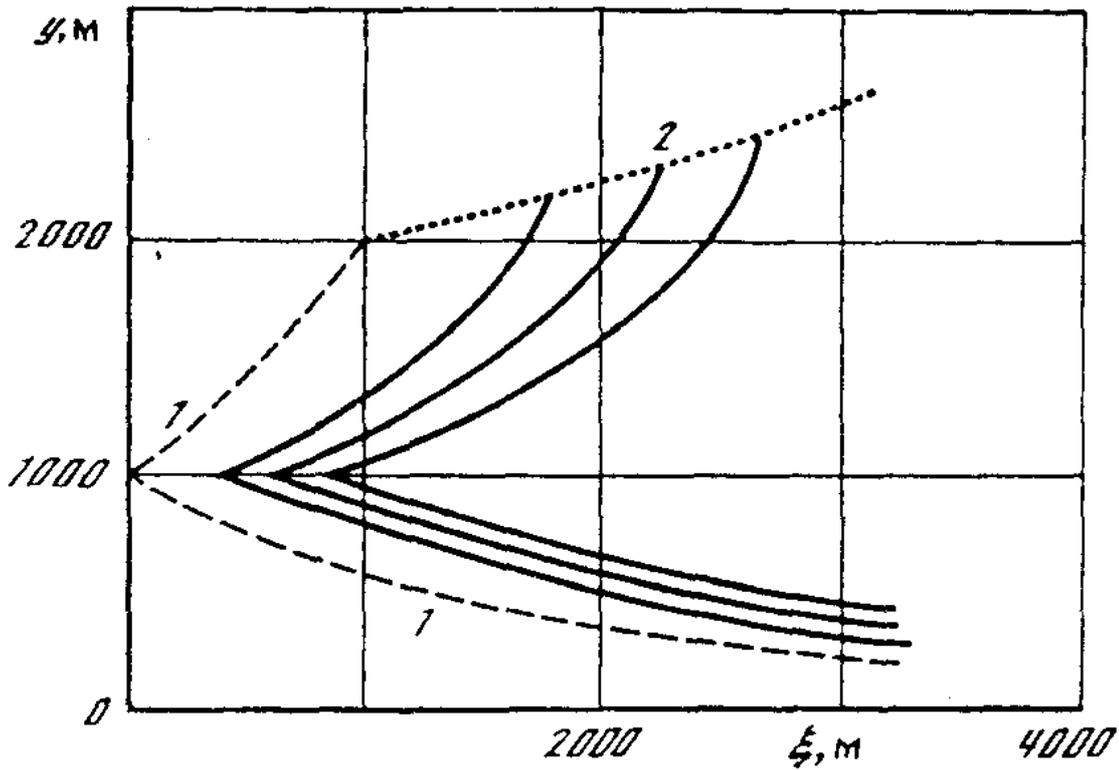

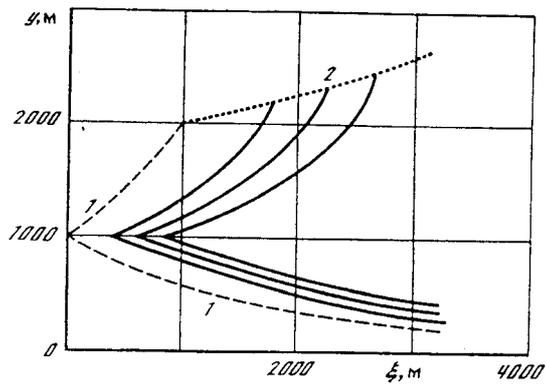

Fig. 1. Phase structure of the wave field in the moving coordinate system. The broken line *1* is the wave front, the continuous lines are phase isolines corresponding to the first three roots of the equation $\Phi'(S)=0$, that is to the first three Fresnel wave antinodes. The dotted line *2* is the envelope of the ray family.